# A New Design of Ultra-Flattened Near-zero Dispersion PCF Using Selectively Liquid Infiltration

Partha Sona Maji* and Partha Roy Chaudhuri

Department of Physics & Meteorology, Indian Institute of Technology Kharagpur-721 302, INDIA
*Tel: +91-3222-283842 Fax: +91-3222-255303,*
*Corresponding author: parthamaji@phy.iitkgp.ernet.in*

**Abstract:** The paper report new results of chromatic dispersion in Photonic Crystal Fibers (PCFs) through appropriate designing of index-guiding triangular-lattice structure devised, with a selective infiltration of only the first air-hole ring with index-matching liquid. Our proposed structure can be implemented for both *ultra-low* and *ultra-flattened* dispersion over a wide wavelength range. The dependence of dispersion parameter of the PCF on infiltrating liquid indices, hole-to-hole distance and air-hole diameter are investigated in details. The result establishes the design to yield a dispersion of 0±0.15ps/ (nm.km) in the communication wavelength band. We propose designs pertaining to infiltrating practical liquid for *near-zero ultra-flat* dispersion of <0±0.48ps/ (nm.km) achievable over a bandwidth of 276-492nm in the wavelength range of 1.26 µm to 1.80µm realization.

## Introduction

The most powerful attributed to photonic crystal fibers (PCFs)[1,2] or microstructure holey fibers is the huge flexibility in the design of transverse geometry by varying the *shape, size* and *positioning* of air-holes in the micro-structured cladding. The hole diameter (*d*) and hole-to-hole spacing (Λ) not only control the dispersion properties, but also the transmission and the nonlinear properties of the fiber as well. Researchers in the past studied in detail this aspect of ultra-flattened dispersion [3-5] over wide wavelength range of interest. Various complicated designs such as different core geometries [6-8] and multiple air-hole diameter in different rings [7-12] have been studied to achieve ultra-flattened dispersion values over wider wavelength bandwidths. However, the realizing technology of complicated structures or PCF having air-holes of different diameters in microstructure cladding remains truly challenging, thus contributing to research of theoretical nature. An alternative route of achieving similar performance is shown to be practicable by filling the air holes with liquid crystals [13-14] or by various liquids such as polymers [15], water [16] and ethanol [17]. Tunable PCG effect and long-period fiber grating has been successfully realized with liquid-filled PCFs [18]. Gundu *et al* [19] designed an ultra-flattened PCF by filling the air holes with selective liquids. With these developments in mind, we revisited the approach of selective hole-filling with liquid towards achieving ultra-flattened dispersion characteristics of PCF over a wide wavelength window. We followed a method reported by Gundu *et al* [19] where the control of dispersion in PCF is accomplished by (i) two air-hole rings infiltrated with liquid with (ii) the precision of refractive indices ($n_L$) of infiltrating liquid required is up-to four decimal and (iii) that for air-hole diameter (*d*) was up-to third decimal to achieve the ultra-flattened nature. With these values for optimized design, practically it is difficult to realize fiber as well as infiltration. Also, this infiltration, if restricted to one air-hole ring, the dispersion behaviour changes drastically. On this understanding, the paper looked for a more realistic dimension and optimization of the PCF geometry and reinvestigated the dispersion effect by exercising the design study through varying the associated parameters. Thus, favouring ease of realization, present research consider the fiber geometry that uses one filled air-hole ring (first ring) and it relies on the values of *d* up-to second decimal such as the precision remains at least up-to 10nm or higher, making the structure resolvable with SEM**.** The values of RI of the infiltrating liquid have been kept up-to third decimal point making the precision practically achievable with the manufacturing companies (*e.g.* M/s Cargille-Sacher Laboratories Inc, USA is having index matching liquid with precision up-to third decimal point).

The selective hole-filling technique provides a couple of advantages. *First*, all the air-holes are the same diameter, which is easier to fabricate compared to fibers with multiple different sub-micron air-hole sizes [7-12]. *Second*, only inner air-hole ring is infiltrated with liquid with certain indices, making it further easier for infiltration of certain liquid point of view. This is why the paper pursued this to select air-hole filling approach for the design of microstructure with the control of target dispersion. Notably, the technique yields are well in designing fibers for various other applications [20-26].

## Dispersion analysis of liquid-filled photonic crystal fiber

Usual conventional PCFs have cladding structures formed by air-holes with the *same* diameter arranged in a *regular* triangular or square lattice. By varying the air-hole diameter (*d*) and hole-to-hole spacing (Λ) of a PCF, the modal properties, in particular, the dispersion properties can be easily engineered. However, the dispersion slope of such PCFs having air-holes of same diameter cannot be tailored in a wide wavelength range. The central idea behind this research is to tailor dispersion closer to zero with a flat slope of the dispersion curve and that too using commonly used regular triangular-lattice structure having air-holes of same size uniformly distributed. A common route of achieving these goals (near-zero and flat dispersion) is by varying the size of air-holes in different layers and is well-known in the literature [7-12]. Because of the fabrication limitation, the concept finds limited use as a practical fiber. The present work looks for the achievement of these targets through a regular conventional PCF by incorporating the effects of filling air-hole ring with a liquid of predetermined refractive index. The paper proposed an index-guiding PCF with the above concept as depicted in Fig. 1. Filling an air-hole with liquid effectively reduces its diameter, depending on the refractive index of the liquid. The fabrication of such a fiber is simplified due to the uniformity of the air-holes in the cladding. To manufacture these PCFs, one must first selectively block specified air-holes and infuse the liquid into the unblocked holes. One possible way is to employ the fusion splicing technique with fusion splicing technique [16-18]. The inner ring of the air-holes can be infiltrated with liquid, first by fusing the outer rings of air-holes with tailored electric arc energies and fusion times [27] and then by immersing one end of the fiber in a liquid reservoir and applying vacuum to the other end of the fiber[15]. This can be a possible way of infiltrating liquid in our case. Another way of selective plug specified air-hole layers in the PCFs is using microscopically position tips with glue [28]. Not only air-hole layers but a single air-hole can be easily blocked by using this technique. In spite of the above methods, one can also selectively infiltrate the liquid into specified air-hole layers from a macroscopic fiber preform to a connected microstructured PCF by using an applied pressure as described in [19]. As we'll see in the optimized air-hole-diameter of 0.40 μm to 0.50 μm, it should be easy to collapse the air holes with diameter 0.40 μm with the method mentioned in [27] and it will be accepted to fill the liquid to the air holes with diameter 0.40 μm, but it will be quite slow and may need the vacuum pump to increase the speed.

There are certain issues related to the infiltration of liquid to the air-holes are, whether the fluid wets glass and how viscous it is. If the liquid does not wet glass then surface tension will oppose entry of the liquid into the hole, making it difficult to fill. One can work out the pressure needed to push such a liquid into a hole given its surface tension and contact angle, and it's likely to require a pressure greater than 1 atmosphere for a 0.40 μm air-hole. In that case, a vacuum pump would be insufficient. If the fluid does wet glass then the hole should fill but the fill speed will depend on viscosity. We can work out how quickly it will fill using the expressions for Poiseuille flow in a pipe. In other words, we can fill the holes (and how quickly), with the given values for surface tension, contact angle and viscosity. With the technology advancing very fast sub-micron filling of air-holes will not be very difficult to achieve.

The design study discussed here consists of a PCF with three rings of air-holes with $C_{6v}$ symmetry with the central air-hole missing as for normal PCF. The inner ring of air-holes is infiltrated with a liquid of certain RI's shown in Fig. 1. By optimizing RI value, $n_L$ of infiltrating liquid combined with PCF geometry namely, pitch (Λ) and air-hole diameter (*d*), ultra-flattened chromatic dispersion can be realized.

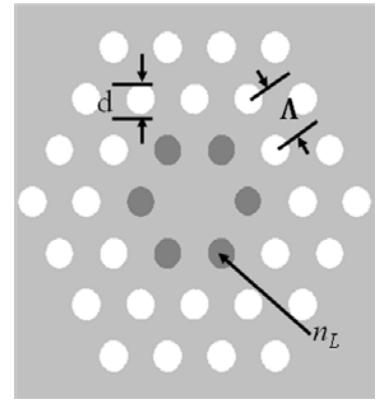

Fig. 1: Cross section of the proposed photonic crystal fiber. The shaded regions represent air holes infiltrated with liquid with refractive indices $n_L$.

We analyze this structure with CUDOS-MOF utilities [29], a Bessel function based software that computes both the real and imaginary refractive indices with certain precision using the multipole method [30-31]. The total dispersion (*D*) is computed with $D = -\dfrac{\lambda}{c}\dfrac{d^2 \text{Re}[n_{eff}]}{d\lambda^2}$ (1).

Here $n_{eff}$ is computed with CUDOS-MOF utilities and *c* is the velocity of light in vacuum.

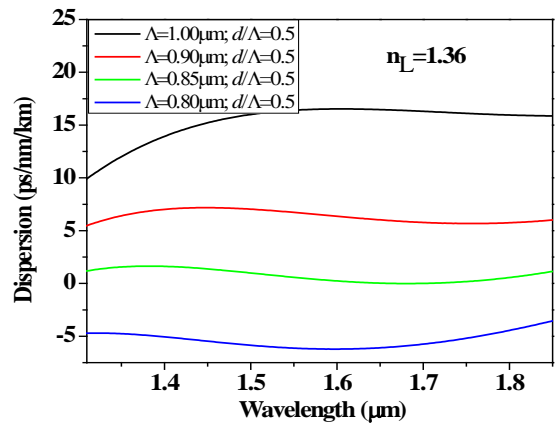

Fig. 2: Computed Dispersion of the PCF as a function of pitch (Λ) keeping $n_L$ and *d* fixed

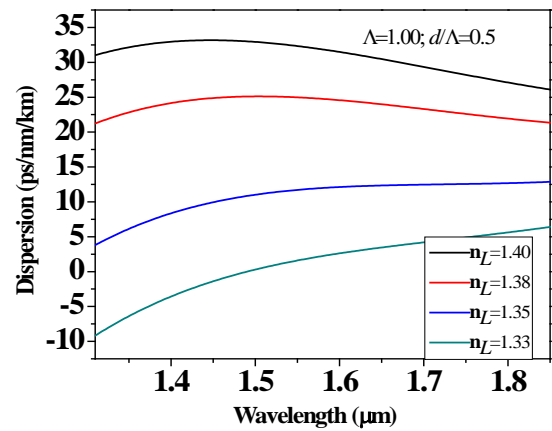

Fig. 3: Dispersion behaviour as calculated for varying $n_L$ values keeping pitch (Λ) and *d* fixed

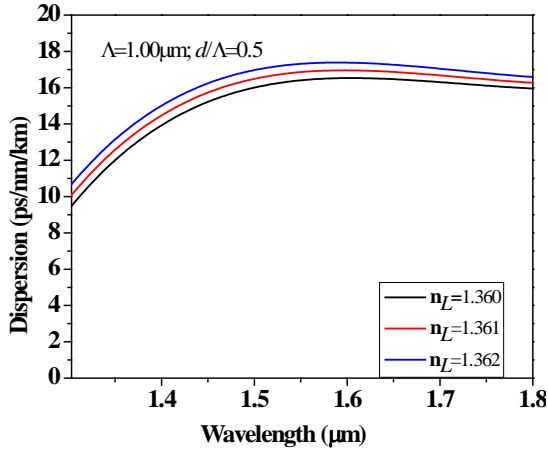

Fig. 4: The sensitivity of *D* for the liquid RI change of 0.001 towards achieving ultra-flat dispersion over a wide wavelength range.

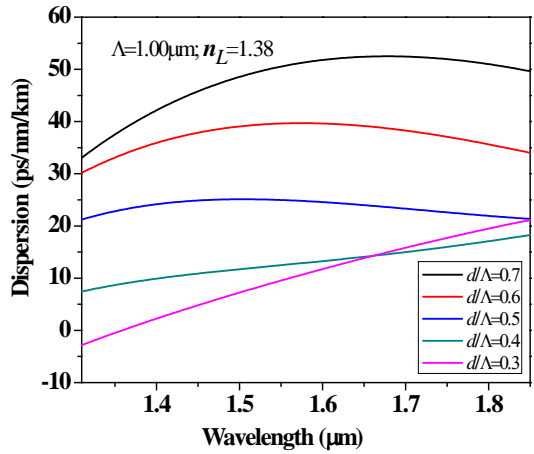

Fig. 5: Variation of Dispersion as a function of air-hole diameter (*d*) when pitch ($\Lambda$) and $n_L$ remain constant.

## Numerical Results towards Optimization for Near Zero Ultra-flattened Dispersion

The approach of the current research optimization relies on varying multi-dimensional parameter space that consists of the liquid RI ($n_L$), the pitch $\Lambda$, and air-hole diameter *(d)* to design ultra flat, near zero dispersion optical fibers. Initially, we consider a liquid that has a constant, wavelength independent refractive index. However, wavelength dependence of the fiber background material (silica glass here) is taken into account, and the refractive index of silica glass is calculated using the Sellmeier formula throughout the study. The atrocious computation of choosing a proper liquid is out of so many available index-matching liquid can be avoided by working initially with an artificial liquid. When a practical liquid is selected, we re-optimize the fiber structure by proper adjustment of $\Lambda$ and *d*. Results with artificial liquid leads us to certain optimization of the parameters. These values give us approximate values of the parameters that we are going to use for practical realization. Now with these values we can select the available liquids and from it we re-adjust the parameters to have ultra-flattened curve. The set of parameters, namely the refractive index of the liquid $n_L$, the pitch $\Lambda$ and the hole diameter *d* are optimized to achieve ultra-flat, near zero dispersion. The procedure is followed in three steps. In the *first* step; we illustrate the effect of varying one of the design parameters on the dispersion curve while the rest are kept constant. This gives us information about the sensitivity of the variation of the parameters value towards the total dispersion. In the *second* stage we took one of the parameters fixed (value obtained from the first step) and optimized the other parameters. Once one parameter is optimized we re-optimize the design by adjusting other parameters. In the *third* stage we select a practical oil (wavelength dependent RI) close to the optimized RI and optimize the other parameters to achieve an ultra-flat near zero dispersion value.

The present section illustrates the *first* stage of the design optimization as follows. Figure 2 shows the effect of $\Lambda$ on the *D* values. The total dispersion changes without much change in its slope for smaller $\Lambda$, while for large $\Lambda$ values the slope increases first and then remains almost flat. From Fig. 3 it can be observed both magnitude and slope of *D* are affected for different values of $n_L$. The graph shows that for lower values of $n_L$, *D* values have always positive slope, whereas for large $n_L$ values *D* increases and then decreases for higher wavelengths. The effect of changing RI up-to third precision has been shown in Fig. 4. The graph shows the sensitivity of *D* for the liquid RI change of 0.001 towards achieving ultra-flat dispersion over a wide wavelength range. This is significant as the thermo-optic coefficient *dn/dT* of the liquids considered here are of the order of $4*10^{-4}/^0C$, limiting the operation within $\pm 3^0$c, but large enough to allow tuning of dispersion by change of temperature [32] .The effect of varying the air-hole diameter *d* is depicted in Fig. 5. It is interesting to observe that for smaller *d* values the slope increases monotonically, where as the slope increases first then decrease for higher *d* values and the slope does not change much for in between *d* values . Thus, the effect of *varying the $\Lambda$ influences the total dispersion*, whereas *d has the desired effect of modifying the dispersion slope*, and *varying $n_L$ modifies both*.

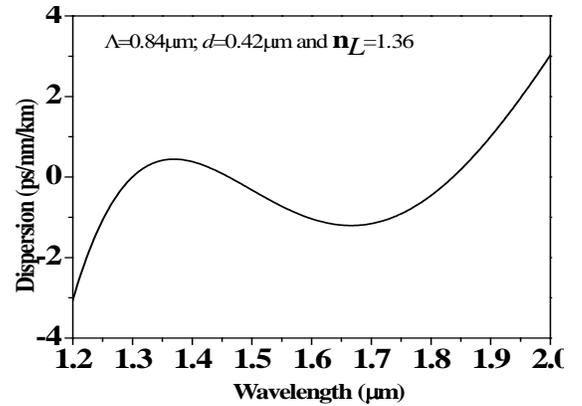

Fig. 6: Ultra-flat dispersion of $0\pm1.20$ps/nm/km over 1245-1910 nm for $\Lambda$=0.84μm, $n_L$=1.36, *d*=0.42μm.

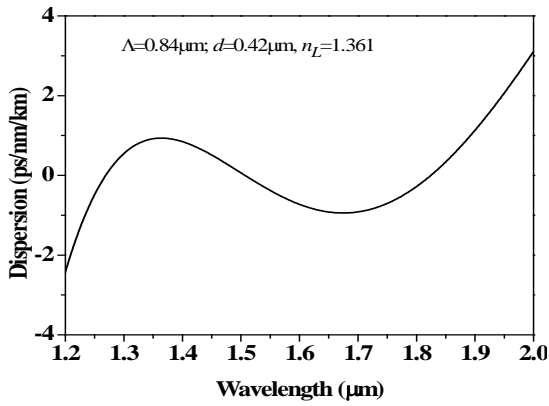

Fig. 7: Ultra-flat dispersion of 0±0.94ps/nm/km over 1236-1888nm for a bandwidth of 652nm for Λ=0.84μm, $n_L$=1.361, $d$=0.42μm.

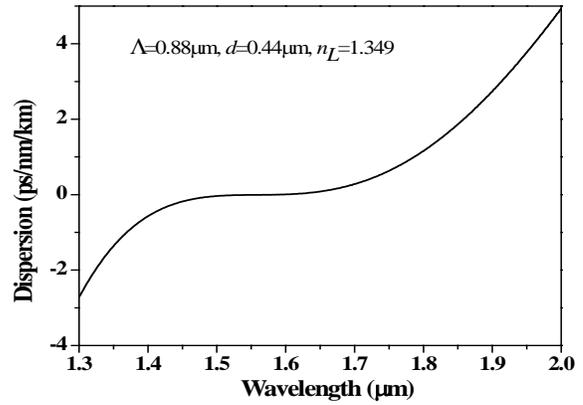

Fig. 9: Ultra-flat "dispersion-less" fiber with a hypothetical liquid of $n_L$=1.349 with the values: Λ=0.88 μm and $d$=0.44 μm.

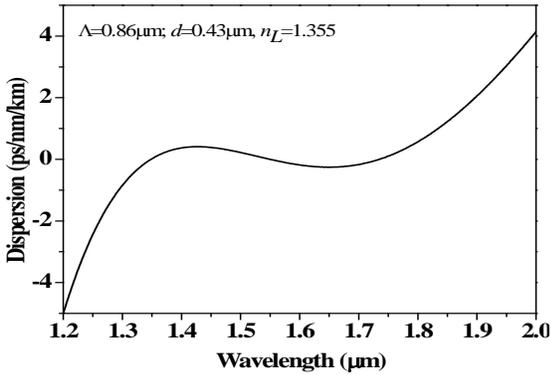

Fig. 8: Ultra-flat dispersion of 0±0.41ps/nm/km over 1322-1784nm for a bandwidth 462nm obtained with the values: Λ=0.86μm, $n_L$=1.355 and $d$=0.43μm.

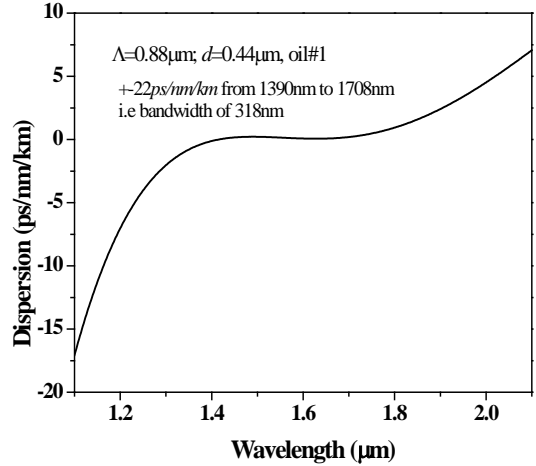

Fig. 10: The ultra-flat dispersion of 0±0.22ps/nm/km over 1390-1708nm with a bandwidth of 630nm obtained with oil#1 with Λ=0.88μm and $d$=0.44μm.

Then, we started the *second* stage of the optimization procedure as the following. Starting with parameters previously considered (*i.e.*, $n_L$=1.36, $d/Λ$=0.5), the value of Λ is varied progressively till we obtain a flat dispersion although not necessarily near zero. Then we successively change $n_L$ and $d$ to either raise or lower the dispersion or to modify its slope. Following the above steps, we obtained the ultra-flattened near zero dispersion in the wavelength region 1245nm to 1910nm *i.e.,* for a bandwidth of 665nm with a tolerance of 0±1.20ps/(nm.km), as shown in Fig. 6 with Λ=0.84μm with $d$=0.42μm and $n_L$=1.36. Noting that $n_L$ precision can be made up-to the third decimal (the available index matching liquid of M/s Cargille-Sacher Laboratories Inc, USA), we obtained an improved result of ultra-flattened *D* values between 0±0.94ps/(nm.km), near zero dispersion point as shown in Fig. 7 in the wavelength range of 1236nm to 1888nm, *i.e.,* with a bandwidth of 652nm. This is achieved with the $n_L$ value of 1.361 keeping Λ=0.84μm with $d$=0.42μm. Two other optimized designs with different parameters are shown in Fig. 8 and Fig. 9 respectively to show the flexibility of the technique. Figure 8 shows an ultra-flattened *D* values between 0±0.41ps/(nm.km), near zero dispersion point in the wavelength range of 1322nm to 1784nm, *i.e.,* with a bandwidth of 462nm with Λ=0.86μm with $d$=0.43μm and $n_L$=1.355. Figure 9 shows an almost dispersion less fiber with Λ=0.88μm with $d$=0.44μm and $n_L$=1.349.

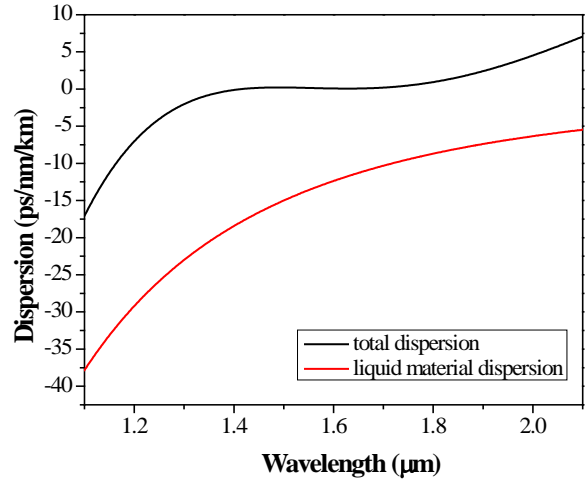

Fig. 11: Contribution of the material dispersion of oil#1 towards the total dispersion for the fiber with Λ=0.88 μm and $d$=0.44 μm. Material dispersion of the liquid contributes significantly towards achieving ultra-flat near zero dispersion.

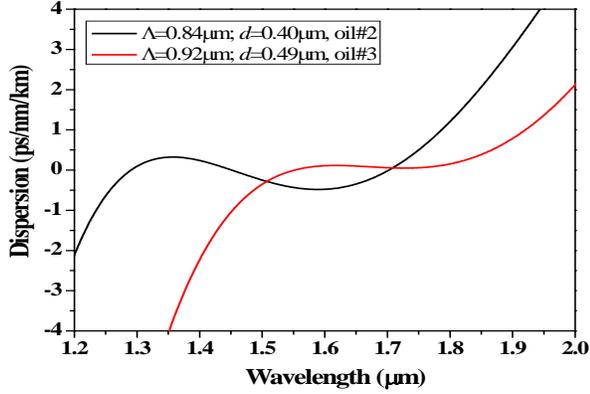

Fig. 12: The dispersion curve obtained with infiltrating the air-hole with oil#2 and oil#3 for an ultra-flat near zero dispersion.

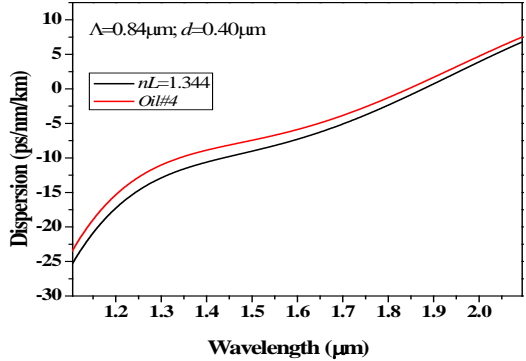

Fig. 13: Comparison of the dispersive properties between PCF infiltrated with *oil#4* and an artificial liquid with RI 1.344.

Having obtained a preliminary design using an artificial, dispersion-less liquid, we continue with the *third* optimization stage. We select an oil (calling it as oil#1) whose RI is nearer to 1.36 in the wavelength range considered and is given by Cauchy equation (2). With this liquid an ultra-flattened between $0\pm0.22 ps/(nm.km)$ near zero D values in the wavelength range 1390nm to 1708nm *i.e.* for a bandwidth of 318nm has been achieved with $\Lambda=0.88\mu m$ with $d=0.44\mu m$ as shown in Fig. 10. Contribution of the material dispersion of the liquid towards the total dispersion has been shown in Fig. 11 for the above structure. The figure clearly shows that the oil has significant contribution towards the total dispersion. The flexibility of the design has been considered with taking in consideration with two other oils (oil#2 and oil#3) that are of different RI than oil#1. The optimized dispersion graphs with these two oils are shown in Fig. 12. Ultra-flattened PCF with *D* values of $0\pm0.48 ps/(nm.km)$ near zero in the wavelength range 1258nm to 1750nm *i.e.* for a bandwidth of 492nm with Oil#2 with $\Lambda=0.84\mu m$ with $d=0.40\mu m$ has been shown in the figure. The ultra-flat near zero *D* value with oil#3 has also been shown in Fig. 12. The flatness is even better with this structure with $\Lambda=0.92\mu m$ with $d=0.49\mu m$ with Oil#3 with *D* values of $0\pm0.15 ps/(nm.km)$ near zero in the wavelength range 1524nm to 1800nm *i.e.* for a bandwidth of 276nm. The optimized parameters along with their dispersion characteristics are summarized in Table 1. These results are new in the design of highly controlled dispersion of PCF. A final study of the comparison of the dispersive properties with an artificial liquid and a practical liquid (we call it as oil#4) are shown in Fig. 13. The RI of oil#4 is governed with Cauchy equation (5) and having RI value of 1.344 around the center of the wavelength range considered and the RI of the artificial liquid is taken to be 1.344. The graph clearly shows that the *D* values changes slightly for the two types of liquids keeping the pattern almost parallel throughout the wavelength range considered.

Cauchy equation of the oils:

Oil#1: $n1(\lambda) = 1.3527514+254675/\lambda^2-1.024360\times10^{11}/\lambda^4$ (2)
Oil#2: $n2(\lambda) = 1.3718235+ 289953/\lambda^2-2.084341\times10^{11}/\lambda^4$ (3)
Oil#3: $n3(\lambda) = 1.3384474+ 228216/\lambda^2-2.293739\times10^{11}/\lambda^4$ (4)
Oil#4: $n4(\lambda) = 1.3432154+ 237036/\lambda^2-4.943692\times10^{10}/\lambda^4$ (5)

where λ are in Angstrom.

Table 1: Summary of the optimized parameters and dispersion properties with three optimized fibers

| Liquids | Optimized parameters | | *D* (*ps/nm/km*) | Wavelength range (nm) | Bandwidth (nm) |
|---|---|---|---|---|---|
| | $d(\mu m)$ | $d(\mu m)$ | | | |
| Oil#1 | 0.88 | 0.44 | 0±0.22 | 1390-1708 | 318 |
| Oil#2 | 0.84 | 0.40 | 0±0.48 | 1258-1750 | 492 |
| Oil#3 | 0.92 | 0.49 | 0±0.15 | 1524-1800 | 276 |

## Conclusions

Towards achieving ultra-low as well as ultra-flattened dispersion in PCF over a wide wavelength window, we have successfully worked out a new structure of selective-liquid filled PCFs. The paper design consists of regular triangular-lattice PCFs having air-holes of same size throughout where the first air-hole ring is infiltrated with liquid of prescribed RI's. Thus, it makes the fabrication realistic using standard technology. With the rapidly advancing technology of microstructure fabrication, we hope it would be possible to realize such a structure. We performed a rigorous series study for optimization of the parameters that yielded an ultra-flat near zero-dispersion PCF with *D* around $0\pm0.48$ *ps/nm/km* in the wavelength range of 1258nm to 1800nm. Three such designs with wavelength dependent liquid have been worked out with dispersion value as small as $0\pm0.15$ *ps/nm/km* has been obtained in the communication wavelength. The paper design will have great influence on many engineering applications, namely dispersion compensation over wide wavelengths, birefringence control, wideband supercontinuum generation, ultra-short soliton pulse propagation and many other photonic device applications like PBG devices and long period fiber gratings.

## Acknowledgment

The authors would like to thank Dr. Boris Kuhlmey, University of Sydney, Australia for providing valuable suggestions in understanding the software for designing and studying the properties of different structures. The authors acknowledge sincerely the Defence Research and Development Organization, Govt. of India and CRF of IIT Kharagpur for the financial support to carry out this research.